\documentclass[preprint,12pt]{elsarticle}

\usepackage{amssymb}
\usepackage{graphicx}

\journal{Solid State Sciences}

\begin{document}

\begin{frontmatter}


\title{The electronic structure of rare-earth iron silicide $R_2$Fe$_3$Si$_5$ superconductors}

\author[INT]{M.J. Winiarski}
\author[INT]{M. Samsel-Czeka{\l}a}
\address[INT]{Institute of Low Temperature and Structure Research, Polish Academy of Sciences, Ok{\'o}lna 2, 50-422 Wroc{\l}aw, Poland}

\begin{abstract}
The electronic structures of $R_2$Fe$_3$Si$_5$ (where $R$ = Lu, Tm, Er, Tb, Yb) intermetallics have been calculated from first principles in local-spin density (LSDA) and LSDA+$U$ approaches. The majority of rare-earth iron silicides, except for the heavy-fermion Yb-based compound, exhibit almost equal values of density of states at the Fermi level (E$_F$) as well as very similar Fermi surface topology. The electronic structure around E$_F$ in the 235-type Fe-based compounds is completely dominated by the Fe 3d states. Thus the different superconducting properties of some members of the $R_2$Fe$_3$Si$_5$ family are rather related to a presence of local magnetic moments of $R$-atoms than to electronic-structure features at E$_F$. 
\end{abstract}


\maketitle

\end{frontmatter}

\section{Introduction}

The ternary rare-earth iron silicides, adopting a tetragonal P4/{\it mnc} (space group no. 128) crystal structure, exhibit a variety of interesting phenomena. The majority of the $R_2$Fe$_3$Si$_5$ (where $R$ = rare-earth atoms) series order antiferromagnetically with the magnetic moments originating only from the lanthanide atoms \cite{Re_1, Re_2, Re_3, Tm_1, Tm_2, Er_1, Yb_1}, except for paramagnetic superconductors with $R=$ Lu, Y, and Sc, reaching superconducting transition temperatures, T$_c$ = 6.25, 1.68, and 4.46 K, respectively \cite{LuYSc_1, LuYSc_2}. A separation of superconducting and antiferromagnetic (AFM) phases occurs in Er$_2$Fe$_3$Si$_5$ \cite{Er_2}. In turn, Tm$_2$Fe$_3$Si$_5$ is a reentrant superconductor under pressure, in which the superconducting phase is destroyed at the N\'{e}el temperature, T$_N$. The properties of this compound are significantly sensitive to any disorder \cite{Tm_3,Tm_4,Tm_5,Tm_6,Tm_7}. Meanwhile, Yb$_2$Fe$_3$Si$_5$ reveals a heavy-fermion (HF) state and Kondo-
lattice behaviour and an absence of superconductivity (SC) \cite{Yb_1}. Any disorder or doping with magnetic impurites, containing f electrons, diminish SC also in Lu$_2$Fe$_3$Si$_5$ \cite{screening_1,screening_2,screening_3}. The overall dependence of magnetic ordering temperatures among $R_2$Fe$_3$Si$_5$ systems \cite{Re_2} suggests that the interplay between separated magnetic and superconducting phases is possible only for heavy lanthanides. In the rest of iron-based 235-type silicides magnetic interactions are too strong for an occurrence of SC.

The specific-heat jumps at T$_c$ in (Lu;Y;Sc)$_2$Fe$_3$Si$_5$ have been successfully described by the two-gap BCS-like model \cite{LuYSc_2, twogap_1, twogap_2, twogap_3, twogap_4, twogap_5}. The inter-band electron scattering between these two weak-coupled gaps opened on the whole Fermi surface (FS) is the reason for an observed anisotropy of superconducting properties in the 235-type iron silicides being rather quasi-2D superconductors \cite{2D_1, 2D_2, 2D_3, 2D_4, 2D_5, 2D_6}. Furthermore, the conventional (phononic) character of an SC mechanism in Lu$_2$Fe$_3$Si$_5$ has been questioned by a fast suppression of T$_c$ caused by non-magnetic impurities and atomic disorder induced by the neutron irradiation \cite{unconv_1, unconv_2, unconv_3}. These effects might be explained by spin-fluctuation mediated scenario of SC, similarly to that postulated in iron (oxy)pnictides.

The two-gap SC with s$\pm$ symmetry is connected with a specific electronic structure, in particular with the FS topology. Former theoretical studies predicted multiband FSs of the 235-type iron superconductors, containing both electronlike and holelike pieces of different dimensionality and orbital character \cite{twogap_1, 235_intermet}, in analogy to the most intensively investigated two-gap superconductor MgB$_2$ \cite{MgB2_FS}. The differences in dimensionality of two FS sheets of this compound appeared to be crucial for its twoband SC phenomenon. In the (Lu;Y;Sc)$_2$Fe$_3$Si$_5$ family, the larger gap may be connected with quasi-one-dimensional (quasi-1D) parts of the holelike FS sheets, while the so-called passive band may contribute to the 3D electronlike FS sheet \cite{2D_6}. Furthermore, in these superconductors, the orbital-dependent contributions of the Fe 3d electrons to the FSs are partially separated from one another \cite{235_intermet}. It enables the possibility of a similar SC mechanism to 
that in iron (oxy)pnictides \cite{orbital}, however, of a much weaker character.

In this work, the electronic structures of other ternary iron silicides (Tm;Er;Tb;Yb)$_2$Fe$_3$Si$_5$ are compared with that in Lu$_2$Fe$_3$Si$_5$, being a reference paramagnetic superconductor. A former study showed that the Fe 3d electrons in this superconductor should play an important role in its SC phenomenon \cite{235_intermet}. However, the substitution of other lanthanide atoms for Lu atomic positions leads to various effects: Tm$_2$Fe$_3$Si$_5$ and Er$_2$Fe$_3$Si$_5$ also exhibit SC, while Tb$_2$Fe$_2$Si$_5$ and Yb$_2$Fe$_2$Si$_5$ are non-superconducting, and in addition the latter is a HF system. Our careful study of electronic structures of these rare-earth iron silicides may help in understanding a variety of physical phenomena exhibited by this family of compounds, especially the examination of their FSs topology from the point of view of two-gap SC and density of states (DOS) at the Fermi level (E$_F$).

\section{Computational details}

Electronic structure calculations for the $R_2$Fe$_3$Si$_5$ (where $R$ = Lu, Tm, Er, Tb, Yb) intermetallics have been performed with the full-potential linearised augmented plane wave (FLAPW) method, implemented in the Wien2k package \cite{Wien2k}. The Perdew-Wang \cite{LDA} form of the local spin-density approximation (LSDA) of an exchange-correlation functional was employed with including the spin-orbit coupling. The LSDA+$U$ approach \cite{LDAU} was used for strongly correlated 4f electrons of Yb, Tm, Er, and Tb atoms. Since a lack of appropriate experimental data for investigated here ternary iron silicides, the value of the effective Coulomb repulsion $U$ of 7 eV was chosen as an average value of those commonly used (and also verified experimentally) to description of the considered $R$-atoms in other compounds, studied in the literature - see e.g. Refs. \cite{LDAU_Er, LDAU_Tm, LDAU_Tm2, LDAU_Yb1, LDAU_Yb2}. For investigated here 235-type compounds, the experimental values of the lattice parameters of 
the unit cells (u.c.) with the $P4/mnc$ symmetry, and atomic positions in each u.c. have been taken from Refs. \cite{Re_1, Tm_2, Er_1, Yb_1, LuYSc_1}. The valence basis cutoff was set to -8.0 eV for a better description of semicore states. Furthermore, the value of R$_{mt}$K$_{max}$ = 8.0 was selected for more accurate plane-wave basis set required by the 4f states coming from the lanthanide atoms. The total energy values were converged with accuracy to 1 meV for the 12$\times$12$\times$12 {\bf k}-point mesh leading to 126 {\bf k}-points in the irreducible part of Brillouin Zone (BZ). The Fermi surface calculations were carried out on more dense mesh of 765 {\bf k}-points in the irreducible part of the BZ. 

\section{Results and discussion}

\begin{figure}
\includegraphics[scale=1.0]{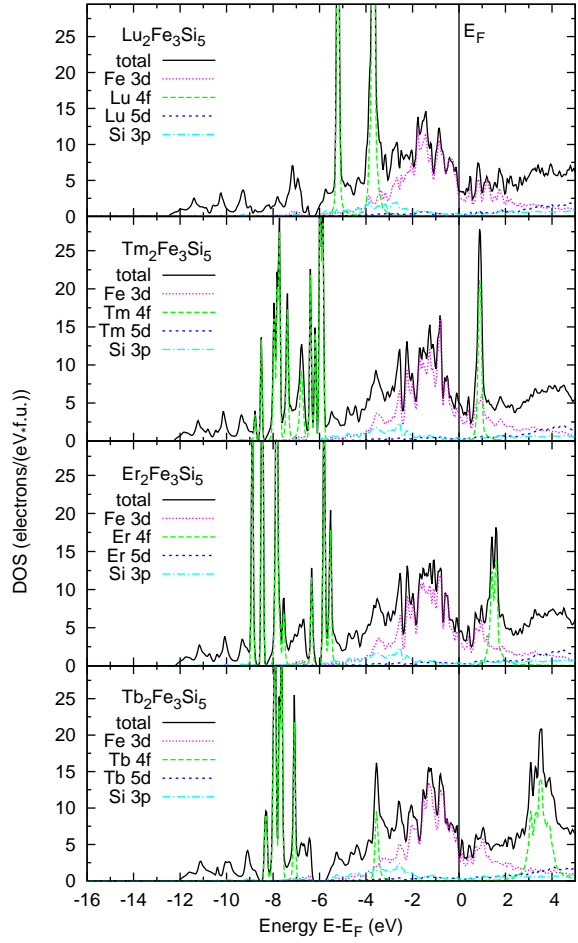}
\caption{Calculated DOS plots for Lu$_2$Fe$_3$Si$_5$ (LSDA) and (Tm;Er;Tb)$_2$Fe$_3$Si$_5$ (LSDA+$U$ with $U$= 7 eV).}
\label{Fig1}
\end{figure}

The DOS plots of $R_2$Fe$_3$Si$_5$ (where $R$ = Lu, Tm, Er, Tb), calculated in either LSDA (for $R$ = Lu) or LSDA+$U$ (for the remaining $Re$ atoms) approaches, are presented in Fig. \ref{Fig1}. It is seen in this figure that the metallic character of the 235-type compounds originates from the wide Fe 3d electron peak centred at about 1.5 eV below E$_F$. The DOS at the Fermi level, N(E$_F$), appears to be completely dominated by this kind of electrons in the whole family of the 235-type systems, similarly to our former results for (Lu;Y;Sc)$_2$Fe$_3$Si$_5$ \cite{235_intermet}. The Si 3p electron orbitals hybridise with the Fe 3d orbitals, predominantly in the energy range of 1-6 eV below E$_F$. Meanwhile, the $R$ 4f electrons are localised at higher binding energy values than 6 eV. At the same time, there are also the 4f peaks located above E$_F$ around 0.9, 1.5, and 2.75 eV for Tm, Er, and Tb atoms, respectively.

The studied here 235-type compounds, except for non-magnetic Lu$_2$Fe$_3$Si$_5$,  exhibit a strong magnetic ordering, due to their local 4f moments, coming from open 4f-electron shells of lanthanide atoms. However, their experimentally observed AFM structures are quite complex \cite{Re_1, Re_2, Re_3, Tm_1, Tm_2, Er_1, Yb_1}. Unfortunately, their non-collinear magnetism can not be investigated with the employed here software, thus the obtained values of the magnetic moments, corresponding to their hypothetical collinear arrangements, are not discussed here. The more so as the main aim of this work is the study of the Fe 3d electrons, carrying negligible magnetic moments in these systems, but the most probably playing a crucial role in the SC phenomenon exhibited in some of these ternaries. As is visible in Fig. \ref{Fig1}, the $R$ 4f electrons only insignificantly influence the DOS at E$_F$.
 
\begin{figure}
\includegraphics[scale=1.0]{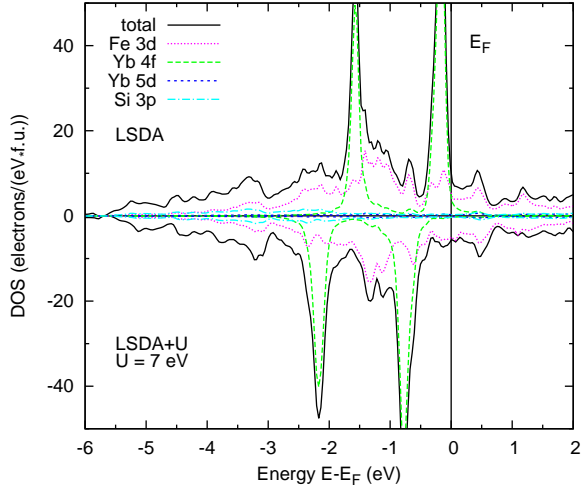}
\caption{DOS plots of Yb$_2$Fe$_3$Si$_5$, computed in LSDA (upper part) and LSDA+$U$ with $U$= 7 eV (lower part) approaches.}
\label{Fig2}
\end{figure}

In turn, for the Yb-based compound, Yb$_2$Fe$_3$Si$_5$, both DOS results (LSDA and LSDA+$U$) are presented in Fig. \ref{Fig2}. Interestingly, in the standard LSDA, the Yb 4f peak is close to E$_F$, contrary to the Lu 4f states that are predominantly localised at higher binding energies. It is known that the 4f-electron shells of free Yb and Lu atoms are completely filled, however, in Yb compounds the 4f electrons may carry magnetic moments and be located close to E$_F$. Since the HF behaviour of Yb$_2$Fe$_3$Si$_5$ can not be described adequately by the standard DFT methods, we have applied both the LSDA and LSDA+$U$ approaches and compared their results in a similar way as it had been done for other heavy-fermion compounds, e.g. YbRh$_2$Si$_2$ \cite{LDAU_Yb1} and YbSi \cite{LDAU_Yb2}. Our results of the usage of $U$ = 7 eV yield a suppression of N(E$_F$) from 10.0 to 5.6 electrons/eV/f.u., which is comparable to other 235-type systems with N(E$_F$) $\approx$ 4-5 electrons/eV/f.u.

\begin{figure}
\includegraphics[scale=1.0]{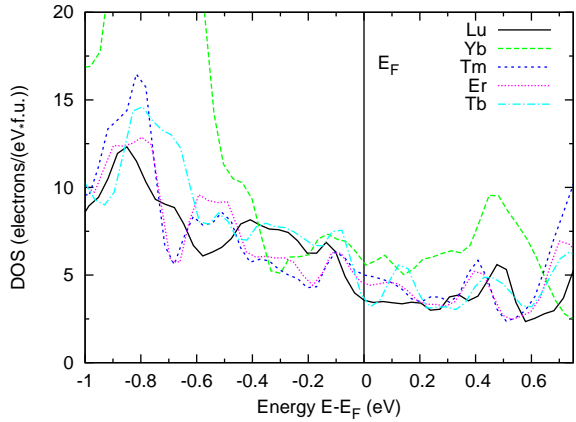}
\caption{DOS plots around E$_F$ for {\it R}$_2$Fe$_3$Si$_5$, where {\it R} = Lu, Tm, Er, Tb, Yb, calculated within LSDA+$U$ (with $U$= 7 eV) approach, except for the case of $R$= Lu, treated by LSDA.}
\label{Fig3}
\end{figure}

The values of N(E$_F$) for all studied here 235-type compounds, collected in Table \ref{table1}, differ insignificantly from one another. Our previous results for isoelectronic (Lu;Y;Sc)$_2$Fe$_3$Si$_5$ showed nearly equal DOSs at E$_F$ as well \cite{235_intermet}. Also the DOSs in a small energy region around E$_F$, depicted in Fig. \ref{Fig3}, are quite similar for almost all studied here compounds, except for the heavy-fermion Yb$_2$Fe$_3$Si$_5$ system. This finding suggests that superconducting properties of ternary iron silicides are rather not related to the DOSs themselves.

\begin{figure}
\includegraphics[scale=0.8]{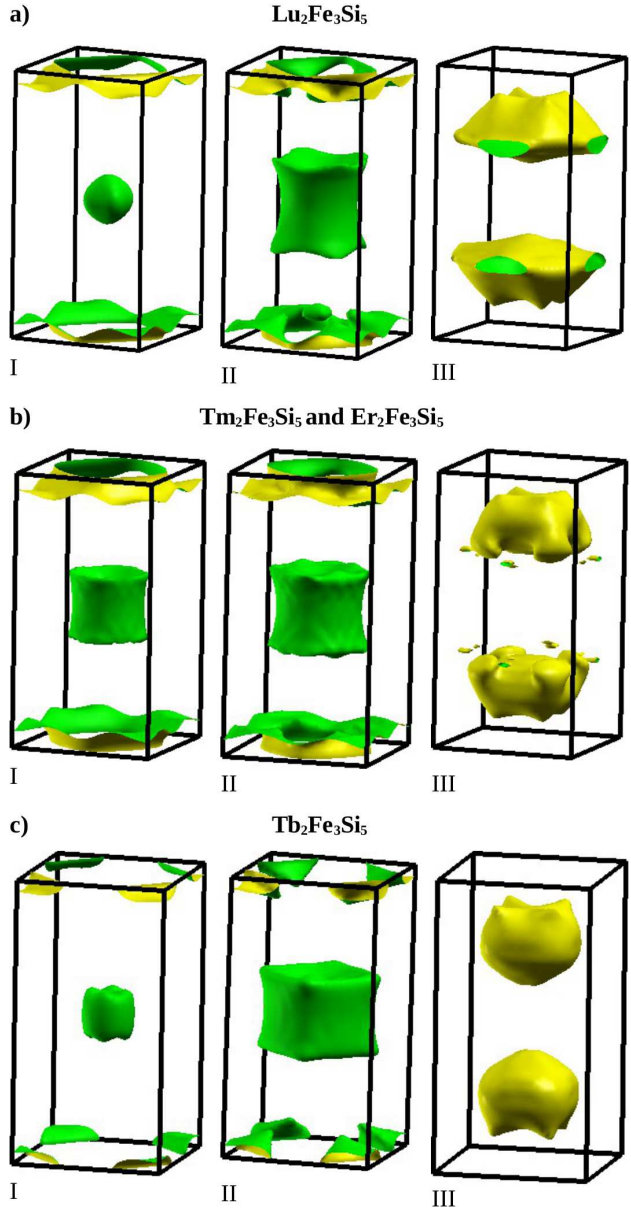}
\caption{Calculated Fermi Surfaces of a) Lu$_2$Fe$_3$Si$_5$ (LSDA) as well as b) (Tm;Er)$_2$Fe$_3$Si$_5$ and c) Tb$_2$Fe$_3$Si$_5$ (LSDA+$U$ with $U$= 7 eV). Fermi surface of each compound consists of three sheets, originating from separate conduction bands (denoted as I-III), drawn within the tetragonal BZ boundaries. Notice that FS sheets of Tm- and Er-based superconductors are almost the same in the scale of this figure, thus only one set of sheets is displayed in part b).}
\label{Fig4}
\end{figure}

The Fermi surfaces of $R_2$Fe$_3$Si$_5$ (where $R$ = Lu, Tm, Er, Tb), computed within either LSDA (for  $R$ = Lu) or LSDA+$U$ (for the other $R$ atoms) approaches, are visualised in Fig. \ref{Fig4}. It is seen that in these compounds, the FSs reveal both holelike and electronlike character of the main three sheets (denoted as I-III) like former results for $R_2$Fe$_3$Si$_5$ (where $R$ = Lu, Y, Sc) superconductors \cite{twogap_1, 235_intermet}. Interestingly, only holelike FS sheets I in Tm- and Er-based compounds distinctly differ as to both the shape and size from that in Lu$_2$Fe$_3$Si$_5$, while the two remaining holelike II and electronlike III sheets of all the considered above compounds are comparable to one another. At the same time, in non-superconducting Tb$_2$Fe$_3$Si$_5$, the central close pocket of FS sheet II  is more cubic from that being rather cylindrical in the other considered here superconductors. Furthermore, FS sheets I i II of Tb$_2$Fe$_3$Si$_5$ contain also some disconnected holelike 
pillows, centred at the corners of the BZ boundaries, instead of large single pieces of reduced dimensionality, being open along the $a$ axis. In this aspect, the FS of Tb$_2$Fe$_3$Si$_5$ resembles rather that reported for Lu$_2$Ru$_3$Si$_5$ \cite{235_intermet}.

\begin{figure}
\includegraphics[scale=0.8]{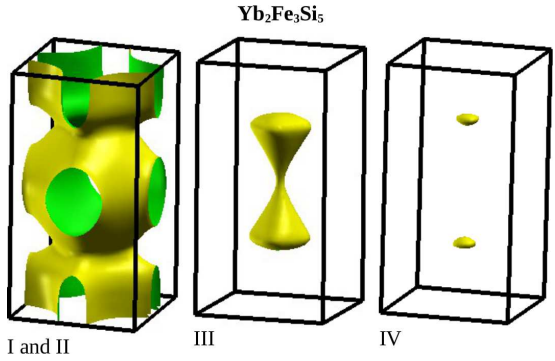}
\caption{Fermi Surface of Yb$_2$Fe$_3$Si$_5$, computed within LSDA+$U$ (with $U$= 7 eV) approach, containing four sheets, denoted as I-IV, drawn within the tetragonal BZ boundaries. Note that FS sheets, originating from two lower bands (I and II) are identical in the scale of this figure, hence, only one of them is displayed.}
\label{Fig5}
\end{figure}

In the iron-based 235-type compounds, the FS shape and dimensionality is determined mainly by the Fe 3d electrons. Only in the heavy-fermion Yb$_2$Fe$_3$Si$_5$, the influence of the Yb 4f electrons on the FS can be significant. It turns out that the FS of this compound, displayed in figure \ref{Fig5}, is completely different from those of the remaining studied here systems, depicted in figure \ref{Fig4}. It is formed by two identical large holelike sheets, I and II, possessing a typically metallic character, which are open along the $a$ and $c$ axes directions, and smaller electronlike sheets, III and IV, in the form of a hourglass and tiny pockets, respectively.

The dimensionality of active and passive superconducting bands for two-gap SC in 235-type Fe-based systems, considered in experimental studies \cite{2D_6}, remains still unclear. Holelike bands are assumed to create the quasi-1D FS pieces, while the 3D electronlike pillows in sheet III, possibly originate from the passive band. It should be mention here that the calculated FSs may be sensitive to structural data. However, for large many-atom u.c.'s of investigated here 235-type compounds, we had to assume only experimental values of all structural parameters, since predicted extremely high costs of their full theoretical optimisation exclude, in practice, a possibility of such calculations. Therefore, the FSs of these systems require further experimental examination by e.g. angle-resolved photoemission spectroscopy (ARPES) to confirm the {\it ab initio} results.

In turn, the analogy between FSs in superconducting Lu$_2$Fe$_3$Si$_5$ and MgB$_2$ is quite interesting. The $\sigma$ (active) and $\pi$ (passive) bands of MgB$_2$ with a distinct orbital character yield also hole- and electronlike FS sheets, respectively \cite{MgB2_FS}. In Lu$_2$Fe$_3$Si$_5$, holelike sheets are related to the Fe 3d $xz$ and $yz$ states while the electronlike sheet is formed clearly by the Fe 3d $z^2$ electrons (see Fig. 3 in Ref. \cite{235_intermet}). In general, the pronounced orbital character may be decisive for multiband SC, particularly in iron (oxy)pnictides \cite{orbital}. Furthermore, the suspected unconventional character of SC in Lu$_2$Fe$_3$Si$_5$ can be universal for all iron-based superconductors.

Electronic structures of ternary iron silicides, except for heavy-fermion Yb$_2$Fe$_3$Si$_5$, are quite similar, suggesting that their SC phenomenon is related to quite subtle electronic features or the suppression of superconducting phase is caused by an occurrence of a magnetic ordering, coming from lanthanide atoms. The disorder-sensitive SC of the 235-type Fe-based superconductors, observed in previous experiments, suggests that further investigations on the influence of nonmagnetic impurities on the electronic structure of Lu$_2$Fe$_3$Si$_5$ might be helpful for better understanding of the SC phenomenon in the family of ternary iron silicides.

\section{Conclusions}

The electronic structures of $R_2$Fe$_3$Si$_5$ (where $R$ = Lu, Tm, Er, Tb, Yb) have been investigated from first principles. Our results indicate that the studied here compounds, except for heavy fermion Yb$_2$Fe$_3$Si$_5$, have a very similar character of both densities of states and the Fermi surfaces, in agreement with our former study on the (Lu;Y;Sc)$_2$Fe$_3$Si$_5$ superconductors \cite{235_intermet}. Moreover, in the 235-type iron-based compounds, the electronic structure around the Fermi level is completely dominated by the Fe 3d electrons. The suppression of superconductivity in Tm$_2$Fe$_3$Si$_5$ and Er$_2$Fe$_3$Si$_5$ and also its complete lack in Tb$_2$Fe$_3$Si$_5$ seem to be related rather to a presence of local magnetic moments than to subtle electronic structure differences, among which a disappearance of quasi-1D character of holelike pieces of FS sheets I and II can also play some role. Presented here results encourage to further investigations of the influence of nonmagnetic 
impurities on electronic structure of the 235-type superconductors that may improve an understanding of the multigap SC phenomenon in iron-based intermetallics.

\section*{Acknowledgements}
This work has been supported by the National Centre for Science in Poland (Grant No. N N202 239540). The calculations were performed in Wroc{\l}aw Centre for Networking and Supercomputing (Project No. 158).

\begin{table}
\caption{Calculated DOS at E$_F$ for {\it R}$_2$Fe$_3$Si$_5$ where {\it R} = Lu, Tm, Er, Tb, Yb atoms.}
\label{table1}
\begin{tabular}{ll}
{\it R} atom &  N(E$_F$) (electrons/eV/f.u.) \\ \hline
Lu & 3.6\\
Tm & 3.8\\
Er & 4.5\\
Tb & 3.8\\
Yb & 5.6\\
\end{tabular}
\end{table}

\end{document}